\documentclass[%
 aip,
 amsmath,amssymb,
preprint,%
]{revtex4-1}

\usepackage{graphicx}
\usepackage{dcolumn}
\usepackage{bm}

\usepackage[utf8]{inputenc}
\usepackage[T1]{fontenc}
\usepackage{mathptmx}
\usepackage{xcolor}
\begin{document}


\title{Nucleation in aqueous NaCl solutions shifts from 1-step to 2-step mechanism on crossing the spinodal}

\author{Hao Jiang}
\affiliation{Department of Chemical and Biomolecular Engineering, University of Pennsylvania, Philadelphia, PA 19104, United States}
\author{Pablo G. Debenedetti}
\author{Athanassios Z. Panagiotopoulos}
\email{azp@princeton.edu.}
\affiliation{ 
Department of Chemical and Biological Engineering, Princeton University, Princeton, NJ 08544, United States 
}%


\date{\today}

\begin{abstract}
In this work, we use large-scale molecular dynamics simulations coupled to free energy calculations to identify for the first time a limit of stability (spinodal) and a change in the nucleation mechanism in aqueous NaCl solutions. This is a system of considerable atmospheric, geological and technical significance. We find that the supersaturated metastable NaCl solution reaches its limit of stability  at sufficiently high salt concentrations, as indicated by the composition dependence of the salt chemical potential, indicating the transition to a phase separation by spinodal decomposition. However, the metastability limit of the NaCl solution does not correspond to spinodal decomposition with respect to crystallization. We find that beyond this spinodal, a liquid/amorphous separation occurs in the aqueous solution, whereby the ions first form disordered  clusters. We term these clusters as ``amorphous salt''. We also identify a transition from one- to two-step crystallization mechanism driven by a spinodal. In particular, crystallization from aqueous NaCl solution beyond the spinodal is a two-step process, in which the ions first phase-separate into disordered amorphous salt clusters, followed by the crystallization of ions in the amorphous salt phase. In contrast, in the aqueous NaCl solution at concentrations lower than the spinodal, crystallization occurs via a one-step process, as the ions aggregate directly into crystalline nuclei. The change of mechanism with increasing supersaturation underscores the importance of an accurate determination of the driving force for phase separation. The study has broader implications on the mechanism for nucleation of crystals from solutions at high supersaturations.
\end{abstract}

\maketitle

\section{Introduction}
Crystallization from solution is a fundamental process that is of interest in many fields, including but not limited to atmospheric sciences, geochemistry and biology.\cite{Erdemir,Sosso, Peckhaus,Driessche} Despite its importance, the nature of the initial, rate-determining and highly non-equilibrium nucleation process, entailing the formation of microscopic ordered precursors (nuclei) of the stable crystal phase, has not been fully revealed at the molecular level, partly due to the fact that the existing experimental techniques lack the spatio-temporal resolution to probe the short-lived nanometer-scale nuclei at the early stages of the nucleation process. Molecular simulations, on the other hand, do not suffer from the lack of temporal or spatial resolution, and the last decade has seen a rapid growth in the number of simulation studies for both homogeneous and heterogeneous nucleation in various systems, ranging from simple models (e.g. Lennard-Jones particles)\cite{Anwar, tenWolde} to realistic systems represented by molecular force fields (e.g. NaCl, CaCO$_3$, urea, etc.)\cite{Haji-Akbari, Li, Soria, Mochizuki, Salvalaglio,Jiang,Lanaro,Mucha,Patey,Tribello,Wallace,Shore}. Many important features of the microscopic mechanisms underlying the nucleation/crystallization of solute molecules or ions from solution have been illustrated. In spite of the rapid progress in understanding the underlying mechanisms using simulations, the accurate calculation of rates remains a challenging task, as the calculation is usually subject to large uncertainties\cite{Zimmermann1}, entails demanding computations,\cite{Jiang} and is often limited to a narrow range of supersaturations.

In general, simulations of nucleation/crystallization from solution are conducted at high solute concentrations, where fast nucleation may be observed within a reasonable amount of simulation time.\cite{Patey, Demichelis, Chakraborty} Despite the invaluable insights provided by many prior simulation studies, a precise determination of the driving force for nucleation, i.e. the solution supersaturation, or more precisely, the difference between chemical potential of the solute in solution and crystal phases, is generally missing in most prior studies. The supersaturation of a simulated solution is often interpreted using the experimental solute solubility,\cite{Mucha,Tribello} however, the actual solubility of the underlying molecular force fields may differ significantly from the experimental value, leading to severe under/overestimation of the driving force. For the few studies that estimated the solubility of the underlying molecular force fields,\cite{Salvalaglio, Salvalaglio1} the chemical potentials of the solute relative to the crystal at different supersaturations were either not known or not calculated with sufficient accuracy. For example, the interpretation of solute chemical potentials in Ref. \citenum{Salvalaglio} was based on the assumption of an ideal solution with unitary activity coefficients. Since the nucleation process is highly sensitive to the driving force, the interpretation of nucleation mechanisms and rates from prior simulations becomes difficult when the supersaturation or the chemical potential of the solution or crystal phases are not properly established. 

The connection between crystallization and liquid-liquid phase separation has been explored previously, both experimentally \cite{Han, Shiomi, Mirsaidov, tenWolde, Vekilov, Asherie, Bonnett, Gower, Burruss, Odom} and by simulations \cite{Demichelis, Wallace}. In protein and colloidal solutions,\cite{tenWolde, Vekilov, Asherie} the liquid-liquid phase transition becomes metastable with respect to the liquid-solid transition when the range of attractive interactions is sufficiently short-ranged. Experiments using polymer blends \cite{Han} that can undergo both liquid-liquid phase separation and crystallization as a function of temperature show an increase in the nucleation rate of crystallization because of the concentration fluctuations caused by spinodal decomposition of two liquids at higher temperatures. Multivalent salts that tend to form hydrated salt phases, such as MgSO${_4}$ and CaCO${_3}$ have been experimentally observed \cite{Burruss, Odom} to form ion-rich liquid phases in supersaturated solutions; a recent study based on molecular dynamics simulations of such solutions \cite{Wallace} proposed a possible mechanism for nucleation proceeding via the formation of ``hydrated clusters''. The study hypothesized the existence of an underlying metastable liquid-liquid transition interfering with nucleation at specific temperatures, but retained the option of crystallizing the solid directly from solution at all concentrations. Two key questions that remain unanswered thus far are (a) how a solution of a salt such as NaCl, that does not form hydrated crystals, loses thermodynamic stability at sufficiently high concentrations and (b) the implications of the loss of stability on the nucleation mechanism. In the current work, we provide unambiguous answers to these questions, by careful simulations of two model systems. Specifically, we study the nucleation in a model Lennard-Jones (LJ) mixture, and of NaCl from supersaturated aqueous solutions. We obtain component chemical potentials that accurately represent the driving force for nucleation, and we identify the stability limit of the supersaturated NaCl solution, for the first time in simulation studies. Prior to reaching the spinodal there are no liquid pre-nucleation clusters and crystal nucleation follows a single-step mechanism. At and beyond the thermodynamic stability limit, we observe a shift from one-step to a two-step nucleation mechanism, i.e. a liquid/amorphous phase separation producing clusters, followed by crystallization, rather than a barrier-free crystal/solution spinodal decomposition. We also calculate crystal nucleation rates at several supersaturations. \\

\section{Methods}

We use the SPC/E (extended simple point charge)\cite{Berendsen} model of water and the Joung-Cheatham\cite{Joung} (JC) NaCl force fields to model supersaturated aqueous NaCl solutions. The SPC/E and JC force fields have been shown to provide reasonable predictions for several solution thermodynamic and transport properties.\cite{Orozco} The JC NaCl model has an equilibrium solubility of 3.7 mol/kg in SPC/E water at 298.15 K and 1 bar, confirmed by both chemical potential\cite{Mester} and direct coexistence methods.\cite{Espinosa} The nucleation rate for this system has been recently studied \cite{Jiang, Jiang2018} using forward flux sampling methods, and from seeding simulations in conjunction with classical nucleation theory (CNT)\cite{Zimmermann,Zimmermann1}; it was demonstrated that at modest supersaturations nucleation follows a classical one-step mechanism.

The difference of chemical potentials of ions between solution and crystal phases is the driving force for nucleation. We calculate the chemical potential of NaCl (ions) in solution from 6.0 to 20.0 mol/kg, following the approach developed by Mester and Panagiotopoulos.\cite{Mester, Mester1} In particular, the chemical potential is estimated in molecular dynamics (MD) simulations in the isothermal-isobaric ensemble as the change in the Gibbs free energy due to the insertion of a pair of ions into the solution. The Gibbs free energy of insertion is obtained through a thermodynamic integration process by slowly switching on the interactions between the inserted pair of ions and the solution. Further details on the calculation of electrolyte chemical potentials can be found in Refs. \citenum{Mester} and \citenum{Mester1}. For all MD simulations associated with the chemical potential calculation, the simulation box has 500 water molecules, and the number of ion pairs ranges from 55 to 180, corresponding to different salt concentrations. Because a relatively small number of ions is used in these calculations of the chemical potential, we do not observe significant crystal nucleus formation during the production stage (around 600 ns) of our simulations. More details on the calculation of electrolyte chemical potentials are provided in the supplementary material. 

In order to follow the progress of nucleation from aqueous solution, it is necessary to distinguish ions that are in the solution and in the crystalline phases. For each ion ($i$), the Steinhardt bond-orientational order parameter\cite{Steinhardt} $q8$ is calculated as,

\begin{equation}
q8(i) = \sqrt{\sum_{l=-8}^{8} |q_{8l}(i)|^2}
\end{equation}
and
\begin{equation}
q_{8l}(i) = \frac{1}{N_B} \sum_{j=0}^{N_B} Y_{8l} \big(\theta (\bm{r_{ij}}), \phi (\bm{r_{ij}}) \big)
\end{equation}
where $Y_{8l}$ are the spherical harmonics, and $\theta (\bm{r_{ij}}) $ and $\phi(\bm{r_{ij}}) $ are the polar and azimuthal angles associated with the vector ($\bm{r_{ij}}$) that connects the central ion ($i$) and one of its neighbor ions ($j$). The summation in Eq. (2) is over the 12 nearest neighbors of the ion $i$ ($N_B$ = 12). We consider an ion to be in the crystalline phase if its $q$8 order parameter is larger than 0.45,  and two crystalline ions that are separated by a distance less than 0.35 nm are considered in the same crystalline nucleus. A similar strategy has been used by Lanaro and Patey\cite{Lanaro} to follow the formation of NaCl nuclei in aqueous solutions. 

When the free energy barrier for nucleation is low (of the order of a few $k_B T$), spontaneous nucleation can be observed from unbiased MD simulations within hundreds of nanoseconds. For such spontaneous nucleation events, the nucleation free energy profile can be extracted from the mean first passage time (MFPT) as,\cite{Wedekind,Wedekind1}
\begin{equation}
\beta G^*(n^*) = \textrm{ln}[B(n)] - \int_{a}^{n^*}\frac{dx'}{B(x')} + C
\end{equation}
and
\begin{equation}
B(x) = \frac{1}{P_{st}(x)} \big[\int_a^x P_{st}(x')dx' - \frac{\tau (x)}{\tau (b)} \big]
\end{equation}
where $\tau (n^*)$ is the MFPT collected from MD simulations as the average time required for the largest nucleus in a trajectory to reach a size of $n^*$ for the first time. $P_{st}(n^*)$ is the steady state probability that a configuration has a largest crystalline nucleus of size $n^*$. $a$ is the boundary of the solution domain, which is taken as the position where $P_{st}$ shows a maximum for a solution before experiencing any crystallization. $b$ is the boundary of the crystal domain, chosen to be 45 in this work, since a critical crystalline nucleus is generally much smaller than 45 for highly supersaturated solutions. 

The above free energy profile from MFPT is expressed as a function of the size of the largest crystalline nucleus. However, the free energy barrier associated with a crystallization process corresponds to the reversible work needed to assemble a crystalline nucleus of size $n$, and it is related to the probability that a system has crystalline nuclei of size $n$ ($N(n)$), rather than the probability of observing the largest nucleus in a system to have size $n$ ($N^*(n)$).\cite{Malek} Thus, the free energy profile of nucleation is expressed as,
\begin{equation}
\beta G(n) = -\textrm{ln} \big[\frac{N(n)}{N(0)}\big]
\end{equation}
where $N(0)$ refers to the average number of ions in the solution phase. The above defined $G(n)$ and $G^*(n^*)$  profiles differ for small $n$ (or $n^*$) and are practically the same for large $n$ (or $n^*$).\cite{Lundrigan} Therefore, following the approach of Lundrigan and Saika-Voivod\cite{Lundrigan}, we first obtain the nucleus size distribution ($N(n)$) from a short MD simulation ($<$ 10 ns) and calculate $G(n)$  using Eq. (5) up to a small $n$ ($<$ 10), as crystallization has not occurred. $G^*(n)$ is then obtained from Eq. (3) and patched onto $G(n)$ for large values of $n$. The nucleation rate $J$ can be estimated by fitting the MFPT ($\tau$) with the following expression,\cite{Lundrigan}
\begin{equation}
\tau(n^*) = \frac{1}{2JV} \bigg[1 + erf\big[c (n^* - n_c) \big] \bigg]
\end{equation}
with $n_c$ (the size of a critical crystalline nucleus) and $c$ considered as fitting parameters. $V$ is the system volume. \\

We also obtained the free energy profile of nucleation using umbrella sampling from hybrid-Monte Carlo (MC) simulations.\cite{Guo_2018} For the umbrella sampling simulations, the size of the largest crystalline nucleus is chosen as the reaction coordinate in the bias potential, expressed as $\phi(n, n_0) = k_0 (n - n_0)^2/2 $. Following Frenkel and Saika-Voivod et al.,\cite{Auer, Sciortino} we perform parallel simulations at different umbrella windows ($n_0$) in order to drive the nucleation. At given $n_0$, the probability that the largest nucleus in the system is of size $n$, defined as $P_{max}(n)$, is calculated, and a histogram of nucleus size distribution in the unbiased ensemble ($N(n)$) is computed from the distribution in the biased ensemble ($\bar{N}(n)$), as $N(n) = <\textrm{exp}[\beta \phi(n, n_0)] \bar{N}(n)>_{biased}$. We then trim $N(n)$ by discarding histogram entries for which $P_{max}(n)$ is less than 0.05 to ensure good sampling statistics. With $N(n)$ thus obtained, the nucleation free energy $G(n)$ is determined in each umbrella sampling window up to an additive constant, and the free energy profile is then constructed by minimizing the difference between overlapping portions of $G(n)$ in each window, as done in Ref. \citenum{Auer} and \citenum{Sciortino}. More details for the umbrella sampling in hybrid-MC simulations are given in the supplementary material. 

\section{Results and discussion}

\subsection{Limit of stability}
\begin{figure}
\centering
\includegraphics[width=0.6\linewidth]{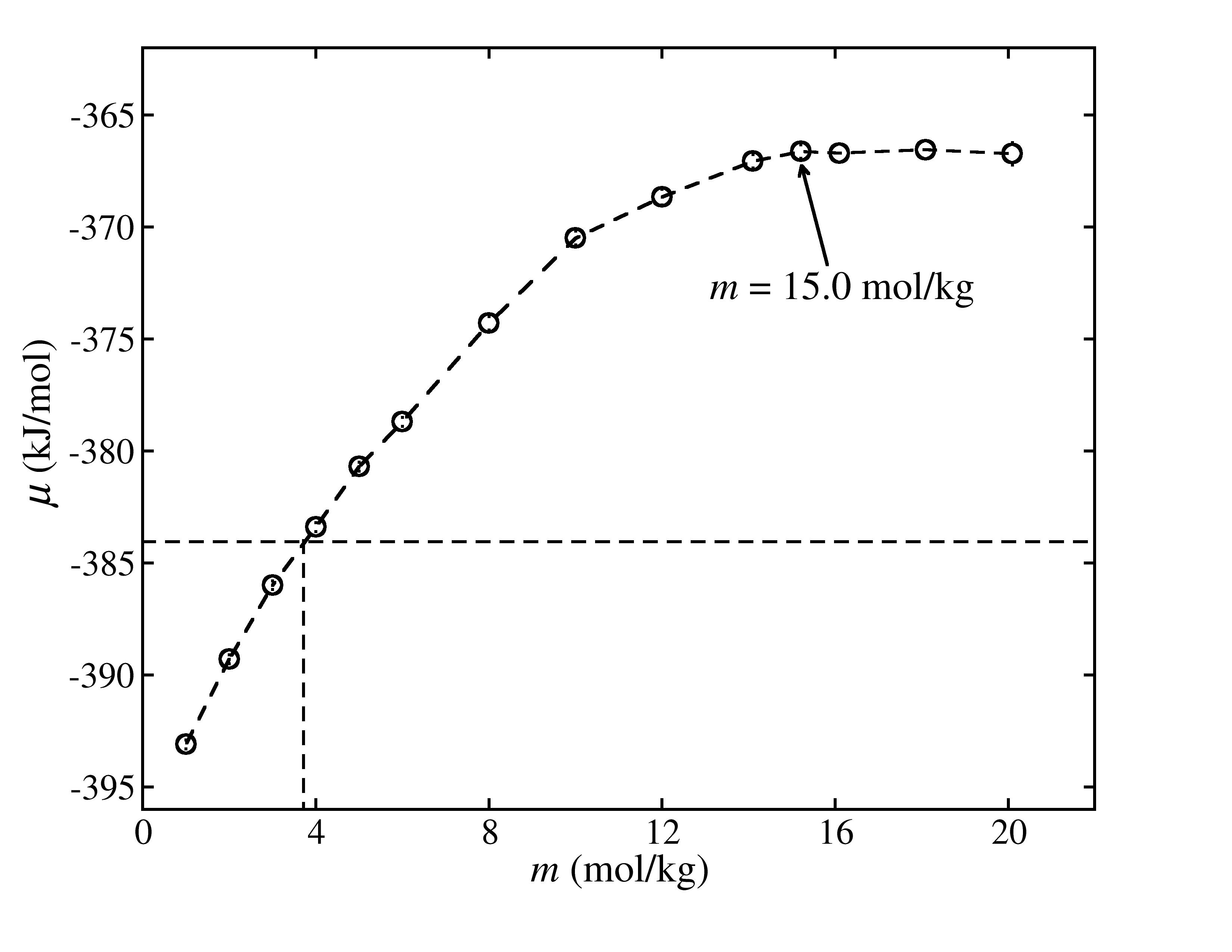}
\caption{Electrolyte chemical potential as a function of salt concentration at 298 K and 1 bar. The curved dashed line is a guide to the eye, and the vertical dashed line indicates the equilibrium solubility for the SPC/E+JC force field combination. The uncertainty of the simulation data is comparable to or smaller than the symbol size. }
\label{fig:mu}
\end{figure}

The chemical potential of NaCl in solution, together with the chemical potential of the crystal, define the driving force for nucleation. The chemical potential of NaCl in solution up to a salt concentration of 20.0 mol/kg is given in Fig \ref{fig:mu}, and the equilibrium solubility of the rock-salt crystal is also shown in the figure as a vertical dashed line. As shown in Fig. \ref{fig:mu}, the electrolyte chemical potential as a function of salt concentration reaches a maximum (-366.6 $\pm$ 0.3 kJ/mol) around 15.0 mol/kg and plateaus beyond this molality, within in simulation uncertainty, up to at least 20 mol/kg (no calculations were performed beyond this salt concentration). Within numerical accuracy of our simulations, the derivative of chemical potential with respect to concentration is zero near 15.0 mol/kg, indicating that the system reaches a thermodynamic stability limit (spinodal) approximately at 15.0 mol/kg. At the spinodal, a new phase must necessarily emerge from the aqueous solution. The constancy of chemical potentials between 15.0 and 20.0 mol/kg suggests that the system experiences a barrier-less phase separation, as the mother phase (solution) is inherently unstable beyond the spinodal. 
Similar to the maximum chemical potential shown in Fig. \ref{fig:mu}, Mou{\v c}ka {\it et al.} identified in their osmotic ensemble Monte Carlo (OEMC) simulations a limiting chemical potential in aqueous NaCl\cite{Mou_ka_2012,Mou_ka_2013} and several other alkali-halide salts solutions\cite{Mou_ka_2012} when the salt concentration is sufficiently high and the solubility predicted by the underlying force field is low. The existence of a limiting chemical potential was found to be a generic feature across different force fields.\cite{Nezbeda_2016} Beyond the salt concentration that corresponds to the limiting chemical potential, precipitate of ions that has rock-salt (FCC) crystalline structure was observed in the supersaturated NaCl solution (see Figure 5 of Ref. \citenum{Mou_ka_2012}). While the observed precipitation phenomenon was connected to the homogeneous crystallization of ions, detailed mechanism associated with such precipitation remains unknown.

\subsection{Crystallization of ions}
\begin{figure}
\centering
\includegraphics[width=0.6\linewidth]{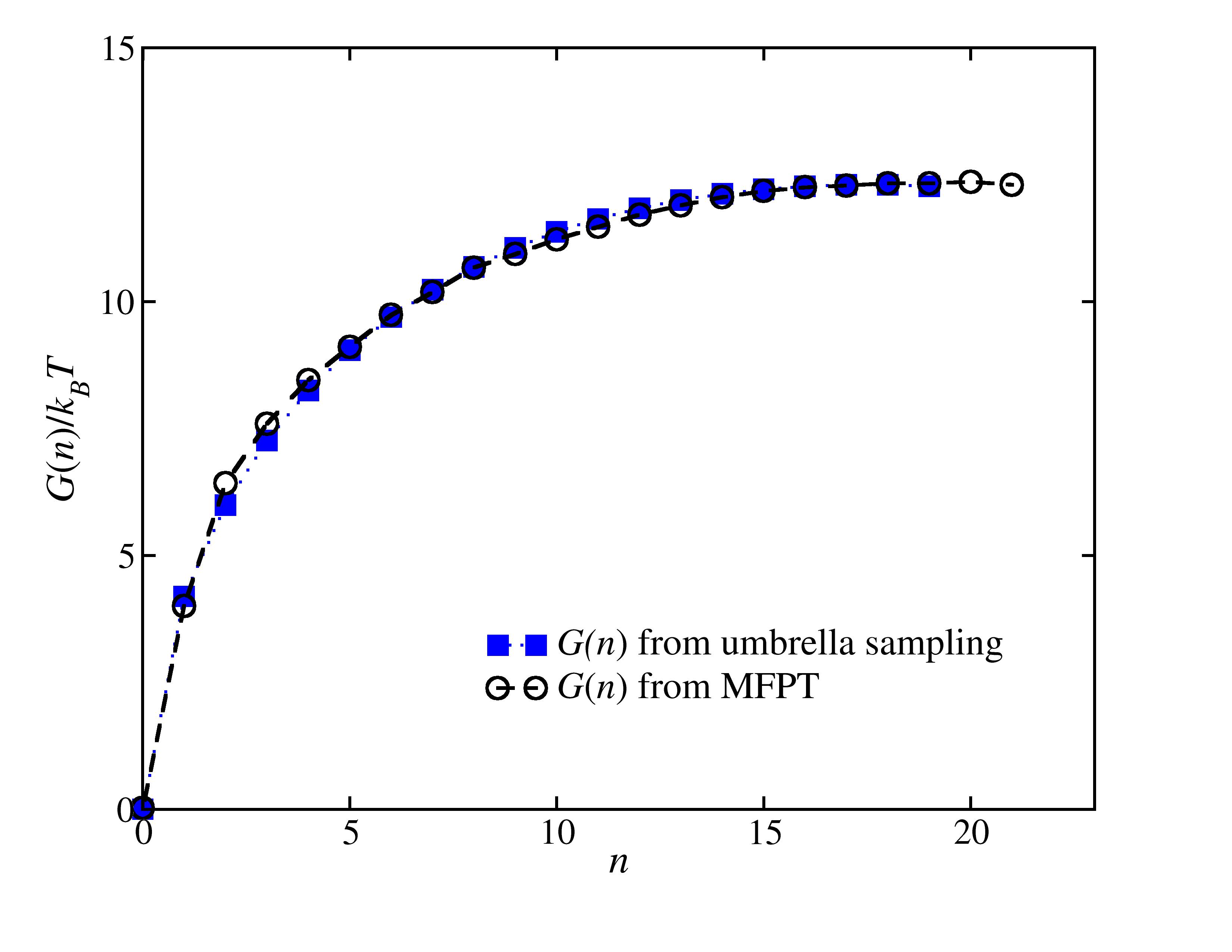}
\caption{Free energy of formation crystalline nuclei of size $n$, $G(n)$, at 15 mol/kg, 298 K and 1 bar.}
\label{fig:delta_G}
\end{figure}

Since the aqueous NaCl solution reaches its stability limit approximately at 15.0 mol/kg, one would normally expect that crystallization of ions from the unstable aqueous solution at and beyond the spinodal is a barrier-free spinodal decomposition process. In order to understand the nucleation mechanism close to the spinodal, we calculated the free energy profile ($G(n)$) for crystallization of ions at 15.0 mol/kg using both the MFPT from MD simulations and umbrella sampling from hybrid-MC simulations, as described above. As shown in Fig. \ref{fig:delta_G}, the free energy profile ($G(n)$) at 15.0 mol/kg has a barrier of approximately 12 $k_BT$, and both the MFPT and umbrella sampling methods give essentially identical results. From umbrella sampling simulations with systems that have 500 and 1000 ions, we obtain the same free energy barriers within simulation uncertainties, indicating that the system size effect on the free energy barrier of nucleation is negligible. If crystallization proceeds by spinodal decomposition, the free energy barrier would vanish. However, this is clearly not the case at the spinodal of the aqueous NaCl solution. 

Given that the system is not experiencing spinodal decomposition towards crystallization, it is interesting to investigate if classical nucleation theory (CNT), which is only valid at low supersaturations, is still applicable at the spinodal for the solution. From CNT, the nucleation rate is calculated as,
\begin{equation}
J_{CNT} = \rho f^+ Z \textrm{exp} (-\frac{-\Delta G}{k_BT})
\end{equation}
where $\rho$ is the number density of NaCl (around 6.5$\times$10$^{27} m^{-3}$ at 15.0 mol/kg). $f^+$ is the attachment rate of ions to a critical nucleus, which can be estimated as the slope of the mean squared evolution of the largest nucleus size, $n^*$, with time ($t$), as $<[n^*(t) - n^*(t=0)]^2>/2t$. $n^*(t=0)$ is close to the critical nucleus size. 100 independent MD simulations are initialized from systems that have largest crystalline nuclei around 20 particles (close to the size of critical crystalline nucleus at 15.0 mol/kg), and the attachment rate is estimated as around 4.6 ns$^{-1}$ (see SI for more details). The Zeldovich factor ($Z$) is associated with the curvature of the free energy profile ($G(n)$) at the barrier and is given by, $Z = \sqrt{G''(n)/(2\pi k_B T)}$. Using the $G(n)$ obtained from umbrella sampling, the Zeldovich factor is calculated as 0.11 $\pm$ 0.03. Considering a nucleation barrier of 12 $k_B T$, the nucleation rate $J$ is calculated from Eq. (5) as 2$\pm 1$$\times$$10^{31}$ $m^{-3} s^{-1}$. Without resorting to CNT, the nucleation rate at 15.0 mol/kg is obtained directly from MD simulations using Eq. (4) from MFPT, as 4$\pm1$$\times 10^{31}$ $m^{-3} s^{-1}$. The rate from CNT agrees very well with the result from the MFPT, within numerical uncertainties. This indicates that CNT is still valid for calculation of the nucleation rates at the spinodal for the prediction of crystal nucleation rate, and again confirms that crystallization at the spinodal has a relatively large (12 k$_B$T) barrier. 

\begin{figure*}
  \centering
  \includegraphics[width=0.6\linewidth]{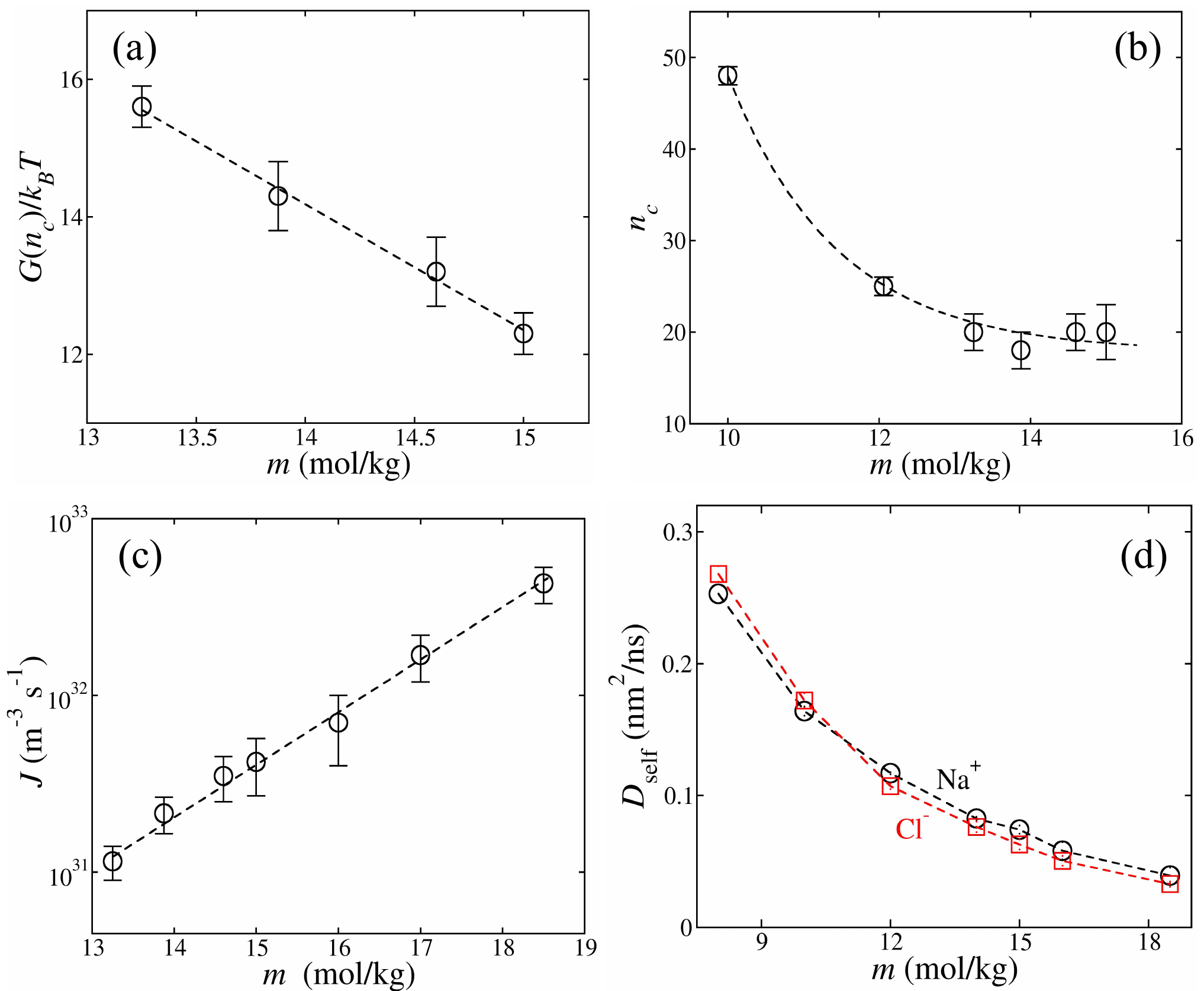} 
  \caption{(a) Nucleation free energy barriers ($G(n_c)$) at different salt concentrations obtained from MFPT and umbrella sampling (at 13.2 and 15.0 mol/kg). (b) Size of the critical nucleus ($n_c$) at different salt concentrations. Data at 10.0 and 12.0 mol/kg are analyzed from data collected during FFS calculation in Ref. \citenum{Jiang} (c) Nucleation rates ($J$) at different salt concentration obtained from MFPT using Eq. (3). (d) Self diffusion coefficient of ions ($D_{self}^{Na^+}$ and $D_{self}^{Cl^-}$ ) at different salt concentrations. The dashed lines are guides to the eye. }
  \label{CNT} 
\end{figure*}

We extended the analysis to a broader range of salt concentrations. Fig. \ref{CNT} shows the nucleation free energy barriers estimated either from MFPT or umbrella sampling, size of critical nuclei estimated from the position of maximum in $G(n)$, and nucleation rates from fitting MFPT with Eq. (4), at different salt concentrations. In addition, the self diffusion coefficients of the ions, estimated from the mean squared displacement using the Einstein relation, are shown in Fig. \ref{CNT}. As expected, the free energy barrier for nucleation of ions into the crystal decreases as the salt concentration increases. The size of critical nucleus ($n_c$) decreases with salt concentration and seems to converge to a constant (around 20), however, the estimation of $n_c$ is subject to large uncertainty for systems close to the spinodal due to the relatively flat free energy profiles (shown in the supplementary material) near the barrier. The nucleation rates ($J$) increase with salt concentration, while the diffusion coefficients of the ions decrease with concentration. If the crystallization of ions took place via spinodal decomposition, which is diffusion controlled, the nucleation rates would decrease with concentration since the diffusion of ions becomes slower as concentration increases. Therefore, the increasing nucleation rates in combination with decreasing diffusion coefficients rules out the possibility of spinodal decomposition-type crystallization. Also, as the system approaches to the spinodal, the free energy barriers and nucleation rates vary smoothly with increasing concentration, without any detectable sudden change. 

\begin{figure*}
  \centering
  \includegraphics[width=0.6\linewidth]{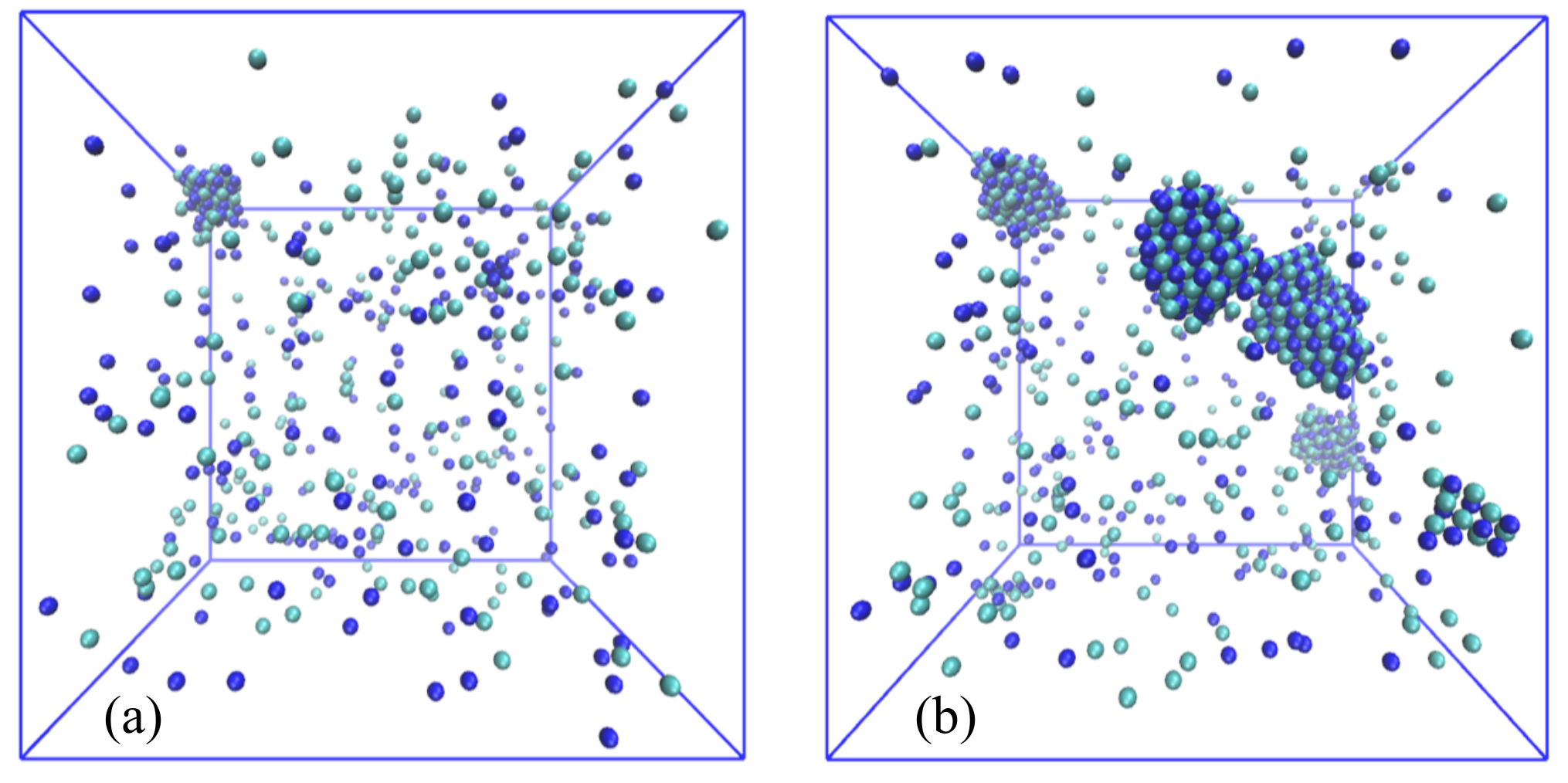} 
  \caption{Snapshots of crystalline nuclei at 15.0 mol/kg. (a) snapshot at 50 ns; (b) snapshot at 150 ns. Only crystalline ions with $q8$ $>$ 0.45 (see Eqs. 1 and 2) are shown in the snapshots. Blue particles: Na$^+$. Cyan particles: Cl$^-$. }
  \label{snapshot_ion} 
\end{figure*}

Fig. \ref{snapshot_ion} shows two snapshots taken at different times from an MD simulation of nucleation at the spinodal (15.0 mol/kg). The system has 7500 ions with a box length around 10.5 nm. Only ions that are labelled as crystalline ($q8$ $>$ 0.45, see Eq. 1) are shown in the snapshots for clarity. Here, we chose to use a system that is much larger than those used in free energy and rates calculations in order to enable a more clear observation of interconnected domains (if they exist), which is a characteristic feature of spinodal decomposition. The snapshots show that at the early stages of nucleation ($t$ = 50 ns), only one small crystalline nucleus emerges from the solution and as the system evolves, more and larger crystalline nuclei are present (e.g. at $t$ = 150 ns), however, the nuclei are not interconnected. The nuclei shows clear rock-salt (FCC) structure with an approximately spherical boundary, which is more consistent with CNT rather than with the Cahn-Hilliard mean-field theory for spinodal decomposition,\cite{Debenedetti} in which the nuclei are ramified with diffuse boundaries. Again, for the crystallization of ions at the spinodal, the system shows characteristics of nucleation/growth instead of spinodal decomposition. Additionally, 3 movies showing the progression of nucleation events at concentrations at 13.8 (before the spinodal), 15.0 (at the spinodal) and 18.5 mol/kg (beyond the spinodal), respectively, are included in the supplementary material. Spinodal decomposition-type crystallization is not observed in these representative movies. 

\subsection{Liquid/amorphous phase separation beyond the spinodal}

The above analysis demonstrates that nucleation/crystallization of ions at the limit of stability does not proceed by spinodal decomposition into a crystal phase. However, the presence of a spinodal indicated by the composition dependence of the chemical potential (Fig. \ref{fig:mu}) certainly suggests that spinodal decomposition occurs in the solution, and a new phase must necessarily emerge from the solution, at and beyond 15 mol/kg. In order to understand this apparent contradiction, we first investigate a binary Lennard-Jones (LJ) system with a LJ solute (S) supersaturated in an LJ solvent (SV), in analogy with the supersaturated NaCl solution.

The interaction parameters of the binary LJ system were taken from Ref. \citenum{Espinosa} (see Table I in Ref. \citenum{Espinosa}), and the mass of Argon is used for both solute and solvent LJ particles. The simulation of the binary LJ mixture was conducted at 50 K and 1 bar, corresponding to a reduced temperature ($k_B$T/$\epsilon$) of 0.44 for the solute (S) and 0.81 for the solvent (SV), respectively, making the binary system exhibit a crystal/solution phase separation at equilibrium, similar to the NaCl/water case. The stable crystal phase in the binary LJ mixture consists of pure solute (S) particles because of the disparity in particle size. The equilibrium solute solubility is ($x_S$=) 0.036(2), as reported in the Ref. \citenum{Espinosa}. 

Following the approach for the calculation of electrolyte chemical potentials, the chemical potential of the solute (S) in the solvent (SV) was obtained from MD simulations at different solute concentrations. The details of these simulations are given in the supplementary material. As shown in Fig \ref{LJ}, similar to the supersaturated aqueous NaCl solution, the binary LJ solution has a spinodal at $x_S$ (mole fraction of solute) approximately to 0.2, indicated by a zero derivative of chemical potential with respect to solute concentration. The chemical potential remains almost a constant from the spinodal to at least $x_S$ = 0.4, suggesting that the system undergoes phase separation by spinodal decomposition.

The free energy barrier for crystallization of the solute particle at $x_S$=0.4, which is beyond the spinodal, was obtained using umbrella sampling from hybrid-MC simulations, as done for the NaCl solution. Details of the hybrid-MC simulations for the binary LJ systems are provided in the supplementary material. We use the procedure proposed by Jungblut and Dellago\cite{Dellago} to identify the crystalline LJ nucleus. In particular, two solute (S) particles are considered as neighbors if they are within 1.5$\sigma$ and the normalized scalar product of their complex $q_{6l}$\cite{Steinhardt} vectors is larger than 0.5. A solute particle that has more than 8 neighbors is considered as a crystalline particle, and two crystalline particles that are within 1.5$\sigma$ of each other belong to the same crystalline nucleus. The free energy profile for crystallization at $x_S$ = 0.4 is given in Fig. \ref{LJ}.  Again, similar to the NaCl solution, the free energy profile of the LJ mixture shows a non-vanishing barrier around 14 $k_B T$ at $x_S$ = 0.4, which is beyond the spinodal. Thus, the crystallization of the solute in the LJ solvent does not proceed by spinodal decomposition, again similar to the NaCl solution. 

However, as shown in the snapshots for systems at two different solute concentrations beyond the spinodal, the solute and solvent LJ particles experience liquid /amorphous phase separation, in which the solute particles aggregate into a metastable disordered phase without crystalline structure. The presence of the solute phase suggests that the spinodal corresponds to a liquid/amorphous phase separation instead of the solution/crystal spinodal decomposition, and the free energy barrier shown in the Fig \ref{LJ} corresponds to the crystallization of solute in the metastable solute amorphous phase. A similar picture for binary LJ solutions has been previously observed by Anwar and Boateng.\cite{Anwar} 

\begin{figure*}
  \centering
  \includegraphics[width=0.6\linewidth]{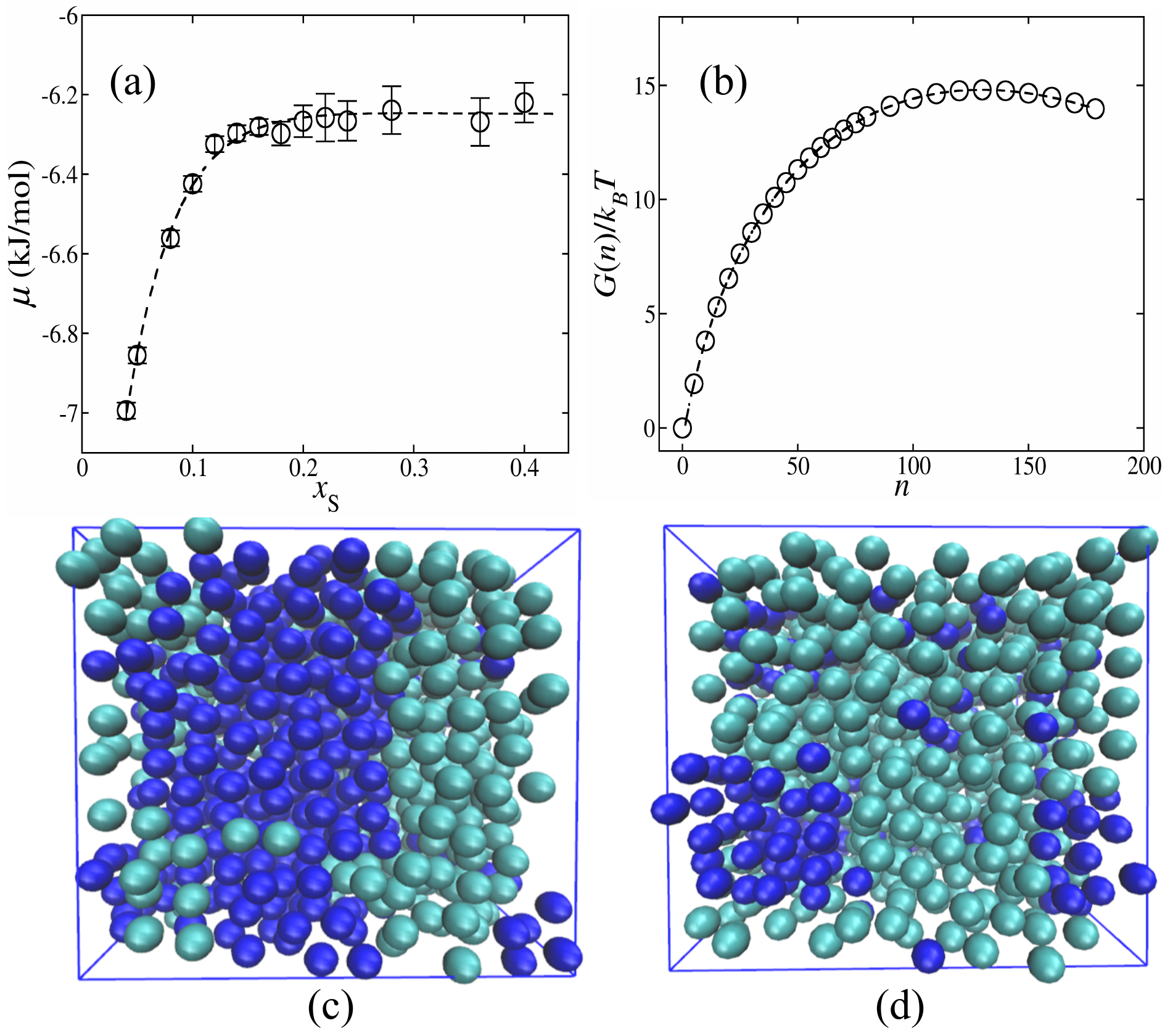} 
  \caption{(a) Chemical potential of the solute LJ component (S) at different mole fractions.  Ideal gas contribution to the chemical potential of solute is set to 0 for a system with solute number density of 10$^{-27}$ m$^3$. (b) Free energy of formation of crystalline solute (S) nuclei of size $n$ at mole fraction of 0.4. (c) Snapshot for the binary LJ system with solute mole fraction ($x_S$) at 0.24, near the spinodal. (d) Snapshot for the binary LJ system with solute mole fraction ($x_S$) at 0.4, beyond the spinodal. The binary mixture has a total of 1000 LJ particles, and both snapshots are taken from NPT-MD simulations at $t$ = 10 ns. Blue particles: solute (S). Cyan particles: solvent (SV). }
  \label{LJ} 
\end{figure*}

\begin{figure*}
  \centering
  \includegraphics[width=0.6\linewidth]{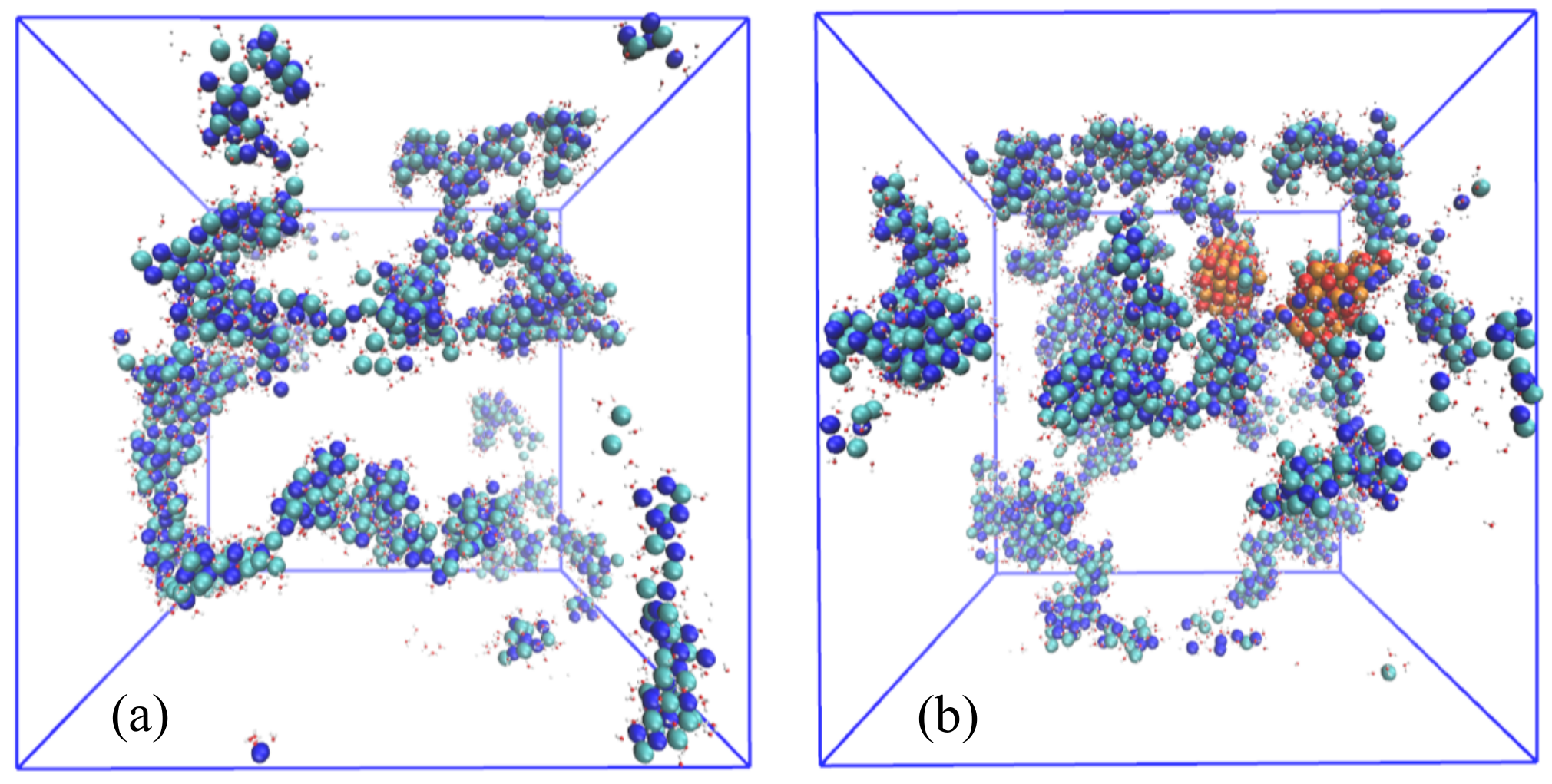} 
  \caption{Snapshots from a MD simulation of aqueous NaCl solution at a concentration of 18.5 mol/kg. (a): snapshot at $t$ = 10 ns; (b): snapshot at $t$ = 70 ns. Only ions with more than 8 ion neighbors (and their hydration water) are shown in the snapshots. Blue and cyan particles: ``amorphous salt'' ions; red and orange particles: crystalline ions.  Small red and white atoms represent the oxygen and hydrogen atoms of water. }
  \label{snapshot_LL} 
\end{figure*}

Based on the behavior of the binary LJ solution, we expect the supersaturated NaCl solution to also exhibit a liquid/amorphous phase separation at and beyond the spinodal. Here, we choose to analyze an MD trajectory for the simulation of an NaCl solution at salt concentration of 18.5 mol/kg, which is beyond the spinodal. Ions that have more than $N_{\rm cut}$ ion neighbors within a radius of $r_{\rm cut}$ are considered to be in an  amorphous phase regardless of the local structure of their neighboring ions, and two amorphous (not solution) ions that are within $r_{\rm cut}$ are considered in the same cluster. We set $r_{\rm cut}$ as 0.45 nm, corresponding roughly to the position of the second peak of the cation/anion pair correlation function. $N_{\rm cut}$ is set to 8, as in the binary LJ mixture. Fig. \ref{snapshot_LL} shows the four largest clusters identified using such criteria, as well as their hydration water molecules surrounding these clusters. A water molecule is considered as a hydration water if it is within 0.4 nm of any ion in a cluster. The snapshots are from a system with 9000 ion pairs and 27000 water molecules, taken at 10 ns (a) and 70 ns (b) of a MD simulation. The two largest crystalline nuclei are also shown in the snapshots. 

As shown in Fig. \ref{snapshot_LL}, the clusters are disordered with dimensions on the order of few nanometers. While the above criteria to identify amorphous ions clusters are ad-hoc, we note that water molecules only surround, but do not penetrate, these relatively large clusters, indicating that the amorphous clusters correspond to a more dense phase compared to the aqueous solution.  At $t$ = 10 ns, which corresponds to the early stages of crystallization, ions form the disordered liquid amorphous clusters, with only a small number of ions (a total of less than 20) forming crystalline nuclei (not shown in the snapshots). Unlike a crystalline nucleus (see Fig. \ref{snapshot_ion}), the amorphous clusters are ramified and have diffuse boundaries, which is consistent with the Cahn-Hilliard mean field theory.\cite{Debenedetti} As the system evolves, at $t$ = 70 ns, part of the ions inside the amorphous salt clusters have already rearranged into crystalline nuclei, shown as red and orange particles.  Similar to the disordered pre-nucleation NaCl clusters here, recent optical Kerr-effect spectroscopy experiments confirmed the presence of large metastable ion clusters in supersaturated sodium thiosulfate solutions.\cite{Reichenbach} 

Based on the above arguments, the spinodal observed from the electrolyte chemical potential vs. concentration curve (Fig. \ref{fig:mu}) identifies the stability limit of the aqueous NaCl solution, with respect to concentration fluctuations, not crystallization, and corresponds to the spontaneous formation of amorphous salt clusters, i.e. a solution/amorphous salt phase separation, rather than the solution/crystal spinodal decomposition. Beyond the spinodal, the system loses its thermodynamic stability as a metastable solution, and starts to form a microscopic amorphous salt phase. Since the amorphous salt clusters have not lost their thermodynamic stability with respect to crystal, the free energy barrier for crystallization is non-vanishing. The NaCl crystal nucleation is thus a two-step process at concentrations beyond the spinodal: a solution/amorphous (liquid/amorphous) phase separation with a vanishing (or low) free energy barrier followed by rearrangement of amorphous ions into crystalline nuclei (liquid/solid phase separation) with a non-vanishing free energy barrier.

Despite the fact that different crystallization processes involve liquid/liquid (or liquid/amorphous) phase separation, the effects of liquid/liquid phase separation on crystallization processes could well be quite varied. In protein and colloid solutions,  the bulk liquid/liquid phase separation that occurs near the fluid/fluid critical point induces large density fluctuations, and the free-energy barrier for crystal nucleation is strongly reduced.\cite{tenWolde} In polymer blends, it has been observed that crystallization may be enhanced by liquid/liquid phase separation, as such phase separation generates phase boundaries or precursors, which subsequently induce heterogeneous crystal nucleation.\cite{Mitra, Gomez}  In contrast to the protein/colloid solutions and polymer blends, the crystallization of NaCl ions evolves smoothly from single-step to two-step, as the nucleation free energy barrier and rates (see Fig. 3) do not change abruptly as the system crosses the spinodal. 

\subsection{Nucleation mechanisms} As discussed above, the nucleation of ions from solution into a crystal at and beyond the spinodal is a two-step process: ions first form disordered amorphous salt clusters, and ions in the amorphous clusters rearrange into crystalline nuclei with a free energy barrier. The two-step (or non-classical) nucleation/crystallization mechanism has been observed, often at large supersaturations, in several systems for crystallization from solution,\cite{Vekilov,Myerson,De} in which the density/concentration fluctuation first leads to the formation of dense phases/precursors, and then solute molecules in the dense phases/precursors transform into an ordered crystal phase with structure fluctuations in a second step. On the other hand, for a single-step nucleation mechanism, concentration and structure fluctuations occur simultaneously, leading directly to the formation of crystalline nuclei.  Several experiments and simulation studies have shown that the single-step nucleation pathway is less favored than the indirect, two-step mechanism for nucleation/crystallization from solutions.\cite{Sleutel} However, there are interesting scenarios where two- and single-step crystallization mechanisms coexist,\cite{Karthika} or a system prefers single/two-step pathway under different conditions. For example, despite many experimental and simulation studies that show a two-(or multi-) step mechanism, crystallization of CaCO$_3$ on organothiol self-assembled monolayer follows a direct one-step pathway without the presence of amorphous precursors prior to nucleation.\cite{Hu} Glucose isomerase, which is known to exhibit characteristics of multi-step nucleation at concentration of 100 mg/mL, crystallizes in solution following a classical one-step mechanism at concentration of 0.1 mg/mL.\cite{Sleutel} The nucleation of urea follows a two-step mechanism in water and acetonitrile solutions, while a single-step pathway is more favored in methanol and ethanol.\cite{Salvalaglio1} 



\begin{figure}
\centering
\includegraphics[width=0.6\linewidth]{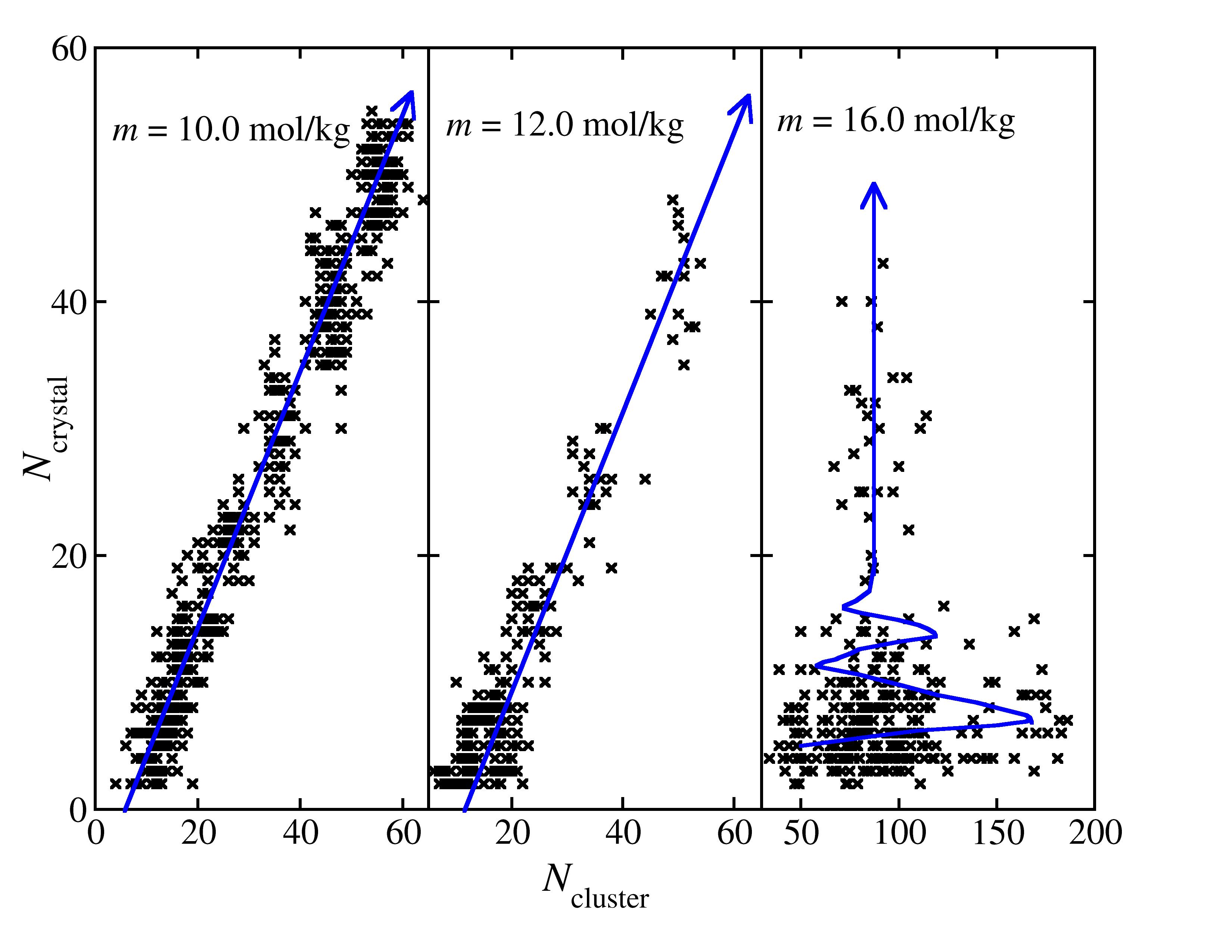}
\caption{Distribution for sizes of the largest ion clusters ($N_{\rm cluster}$) and sizes of the largest crystalline nuclei ($N_{\rm crystal}$) at salt concentrations of 10.0, 12.0 and 16.0 mol/kg. Solid lines indicate the evolution of $N_{\rm cluster}$ and $N_{\rm crystal}$ with time.}
\label{fig:Ncry}
\end{figure}

Since the nucleation of NaCl crystals follows a two-step process at and beyond the spinodal, it would be interesting to investigate if the two-step mechanism also holds before the system reaches its spinodal. Here, we analyze trajectories collected from unbiased MD simulations that include nucleation events, for systems at different salt concentrations. Fig. \ref{fig:Ncry} shows the distributions of sizes of largest amorphous salt clusters ($N_{\rm cluster}$) versus the sizes of largest crystalline nuclei ($N_{\rm crystal}$) for NaCl solutions before reaching the spinodal (at concentrations of 10.0 and 12.0 mol/kg) and after the spinodal (at 16.0 mol/kg). Each symbol in Fig. \ref{fig:Ncry} corresponds to a configuration collected during MD simulations. The data at 10.0 mol/kg and 12.0 mol/kg are based on configurations collected at milestones of forward flux sampling (FFS) calculations of nucleation rates in Ref. \citenum{Jiang}. The data at 16.0 mol/kg are obtained using configurations sampled from an unbiased MD simulation (with 1000 ions pairs) of 120 ns, which produces a crystalline nucleus of 45 ions. Crystalline nuclei that are very small (less than 3-4 ions) and do not reside in the largest disordered clusters are excluded from the analysis. It is noted that the configurations collected at different FFS milestones are joined by stochastic events due to the randomization of particles velocities required by FFS algorithm, while the configurations collected for system at 16.0 mol/kg belong to a deterministic MD trajectory. We do not expect such difference in sampling methods to affect $N_{\rm cluster}$ and $N_{\rm crystal}$, as these quantities convey structural information. The distribution of sizes of the largest amorphous salt clusters ($N_{\rm cluster}$) quantifies concentration fluctuations, while the distribution of sizes of the largest crystalline nuclei ($N_{\rm crystal}$) measures structural fluctuations. As shown in  Fig. \ref{fig:Ncry}, for solutions before reaching the spinodal (10.0 and 12.0 mol/kg), $N_{\rm cluster}$ correlates linearly with $N_{\rm crystal}$, which indicates a simultaneous fluctuation of concentration and structure, consistent with a single-step nucleation mechanism. In fact, at low supersaturations away from the spinodal, the largest amorphous salt cluster, identified using the criteria discussed above, is essentially the ordered largest crystalline nucleus with minor differences associated with the identification of ions at the interface of the cluster/nucleus. As nucleation proceeds, the ions nucleate into a crystalline nucleus simultaneously as they aggregate. After the system passes the spinodal (at 16.0 mol/kg), $N_{\rm cluster}$ has a much wider distribution at early stages of nucleation (small $N_{\rm crystal}$), and the concentration fluctuations are not coupled with structure fluctuations, which indicates a two-step nucleation mechanism. 

The above analysis has shown that there is a transition from single to two-step crystallization mechanism in the supersaturated NaCl solution, and such transition is driven by the microscopic solution/amorphous salt spinodal. It is possible that the two-step nucleation is a mechanism that is only associated with systems that are close to, or beyond, their stability limits, and particles nucleate via the single-step mechanism for systems well before reaching the spinodal. The general belief that classical nucleation theory is only strictly applicable at low supersaturations may be interpreted more precisely as suggesting that the theory is only strictly valid at supersaturations well before reaching the solution spinodal.  Such hypothesis may possibly explain the preference of single/two-step crystallization pathway under different solution conditions. Regardless the validity of the proposed hypothesis, the spinodal driven transition of crystallization mechanisms suggests that an accurate determination of crystallization driving forces from the calculation of solution/crystal chemical potentials is of special importance in order to interpret the nucleation mechanism from simulations.


\section{Conclusions}
In this work, we have examined the nucleation of ions from aqueous NaCl solutions using molecular simulations. For the SPC/E water and JC ions force fields, the chemical potential of the electrolyte starts to plateau at salt concentrations around 15.0 mol/kg, suggesting the presence of a spinodal. The spinodal indicates that the aqueous NaCl solution reaches its limit of stability with respect to concentration fluctuations, and a phase separation through spinodal decomposition occurs in the solution. However, the nucleation of ions into rock-salt crystal at the spinodal shows no signatures of  solution-crystal spinodal decomposition: the free energy barrier to crystallization does not vanish, the critical crystalline nuclei remains finite and have an approximately spherical shape, and the nucleation rates increase with ion concentration, in spite of the corresponding slowing down of ion diffusive motion. Therefore, the spinodal in the system is not associated with crystal/solution phase separation. It is found that the solution at the spinodal experiences a microscopic liquid/amorphous phase separation similar to that previously observed in a binary Lennard-Jones solution \cite{Anwar}. The ions form a metastable amorphous salt phase in transient coexistence with the solution, and the nucleation of ions occurs in the amorphous salt clusters with a free energy barrier. While the nucleation is a two-step process beyond the spinodal, a single-step nucleation to the crystalline phase is observed for solutions before reaching the spinodal.  The transition from one-step to two-step crystallization mechanism, driven by the solution spinodal, suggests that an accurate knowledge of solution and crystal chemical potentials is of importance in order to obtain a proper interpretation of mechanisms from simulations of nucleation.  We hypothesize that the two-step mechanism observed in NaCl aqueous solutions and in other systems with different solute/solvent interactions (e.g. binary LJ system) may be a general feature for systems in the vicinity of a stability limit. 

\section{Supplementary Material}
See supplementary materials for details of molecular dynamics and hybrid Monte Carlo simulations, mean first passage time for estimation of nucleation rates, crystallization free energy profile at salt concentrations of 13.2, 13.8, 14.6, and 15.0 mol/kg, attachment rates of critical nuclei at 15.0 mol/kg, and movies showing progression of nucleation events at 13.8, 15.0 and 18.5 mol/kg.

\section{Acknowledgements}
Financial support for this work was provided by the Office of Basic Energy Sciences, U.S. Department of Energy,
under Award No. DE-SC0002128. Additional support was provided by the National Oceanic and Atmospheric Administration (Cooperative Institute for Climate Science Award No. AWD 1004131) and the U.S. National Science Foundation under award CBET-1402166.

\bibliography{ms}

\end{document}